# Impact of Vehicular Communications Security on Transportation Safety

Panos Papadimitratos*, Giorgio Calandriello†, Jean-Pierre Hubaux* and Antonio Lioy†
*Laboratory for Computer Communications and Applications
EPFL, Switzerland
Email: {panos.papadimitratos, jean-pierre.hubaux}@epfl.ch
†Dipartimento di Automatica e Informatica
Politecnico di Torino, Italy
Email: {giorgio.calandriello, lioy}@polito.it

*Abstract*—Transportation safety, one of the main driving forces of the development of vehicular communication (VC) systems, relies on high-rate safety messaging (beaconing). At the same time, there is consensus among authorities, industry, and academia on the need to secure VC systems. With specific proposals in the literature, a critical question must be answered: can secure VC systems be practical and satisfy the requirements of safety applications, in spite of the significant communication and processing overhead and other restrictions security and privacy-enhancing mechanisms impose? To answer this question, we investigate in this paper the following three dimensions for secure and privacy-enhancing VC schemes: the reliability of communication, the processing overhead at each node, and the impact on a safety application. The results indicate that with the appropriate system design, including sufficiently high processing power, applications enabled by secure VC can be in practice as effective as those enabled by unsecured VC.

## I. INTRODUCTION

Vehicular communication (VC) systems are developed as a means to enhance transportation safety and efficiency. Vehicles and road-side infrastructure units (RSUs) are equipped with on-board sensors, computers, and wireless transceivers. Vehicle-to-vehicle (V2V) and vehicle-to-infrastructure (V2I) communication enable primarily safety applications. Many research and development projects, including the Car-to-Car Communication Consortium in Europe and the US Department of Transportation VII initiative, converge towards a design with vehicles frequently *beaconing* their position along with warnings on their condition or the environment. Typical beaconing periods considered are in the order of one beacon per 100 milliseconds per vehicle.

At the same time, it has been understood that VC systems are vulnerable to attacks and that the privacy of their users is at stake. For example, an attacker could inject messages with false information, or collect vehicle messages to track their locations and infer sensitive user data. As a result, the research community in industry and academia, with the endorsement of authorities, has undertaken three major efforts to design security and privacy enhancing solutions for VC: the NoW project [8], the IEEE 1609.2 working group [9], and the SeVeCom project [11].

A few basic ideas transcend all these efforts to develop VC security architectures. They all build on top of a currently well-understood vehicular communication protocol stack that includes safety beaconing. Moreover, they all utilize a *Certification Authority (CA)* and public key cryptography to protect V2V and V2I messages. Their primary requirements are message authentication, integrity, and non-repudiation, as well as protection of private user information. To address those apparently contradictory goals, they all rely on the concept of *pseudonymity* or *pseudonymous authentication* [8], [9], [11]: they require that (i) each vehicle (node) is equipped with multiple certified public keys (pseudonyms) that do not reveal the node identity, and (ii) the vehicle uses them alternately, each for a short period of time, so that messages signed under different pseudonyms cannot be linked.

It is becoming clear that the security overhead of such systems will be significant; for example, each safety beacon has to be signed, and each vehicle has to validate every 100 milliseconds beacons from several dozens of vehicles within range. Which, not to forget, may essentially change their identity (pseudonym) at any point in time, thus making it harder for their neighboring vehicles to validate their signed beacons.

The immediate question, for designers and users of vehicular communication systems alike, arises: What is the effect of security and notably these broadly accepted pseudonym-based mechanisms on safety applications? In plain terms, can, for example, vehicle collisions still be avoided when an emergency braking situation arises?

There has been a timid approach to answer this question in our earlier work [5], whose main contribution was the simplification of key (pseudonym) management while satisfying privacy and security requirements. But, in terms of evaluating the impact of security on the VC system effectiveness, we analyzed only a simple transportation scenario: the distance at which a fast-approaching vehicle receives an emergency braking signal from a vehicle ahead.

In this paper, we extend the work in [5] to investigate the impact of security on transportation safety. First, we consider more realistic and complex scenarios, in particular, a platoon of one hundred cars. We develop a simulation environment and provide an evaluation that takes into consideration factors critical for vehicle safety. Primarily, the probability of (safety) message reception, especially as the channel load may change, and the processing load. We study the impact of an emergency situation throughout the platoon: we measure the occurrences of vehicle collisions for different environments and different security and privacy protection schemes, and compare against the effectiveness of a basic "no security" safety beaconing scheme. We emphasize that our objective here is not to propose new security and privacy enhancing schemes. Rather, we evaluate the state-of-the-art approaches and seek to identify if and how they can be practical, satisfying the requirements of safety applications.

In the rest of the paper, Sec. II outlines the secure communication schemes we consider in our evaluation. We present the simulation setup for our analysis in Sec. III. Our results, for each of the considered schemes, follow, on three fronts: (i) the reliability of communication (Sec. IV), (ii) the processing load at each node (Sec. V), and (iii) the overall impact on the transportation safety, expressed as the proportion of collided vehicles, under various conditions (Sec. VI). Even under high-stress network scenarios, our findings are encouraging: with appropriate system design, the achieved safety levels for secure and privacy-enhancing VC systems can be practically indistinguishable from those achieved without security.

## II. SECURE AND PRIVACY-ENHANCING VEHICULAR COMMUNICATION

### A. Baseline Pseudonyms (BP)

Following the notation in [5], each vehicle (node) $V$ has a set of *pseudonyms*, i.e., public keys certified by the CA without any information that identifies $V$. If $K_V^i$ is $V$'s $i$-th pseudonym, the certificate $Cert_{CA}(K_V^i)$ is simply the CA signature on $K_V^i$. The node uses the private key $k_V^i$ corresponding to the pseudonym $K_V^i$ to sign messages. The signer's pseudonym and certificate are attached in each message, with $\sigma_{k_V^i}(m)$ the signature on message $m$ under $K_V^i$. Each pseudonym is used at most for a period of $\tau$ seconds and it is then discarded. When a signed message is received, nodes validate, with the public key of the CA, $Cert_{CA}(K_V^i)$, and then $\sigma_{k_V^i}(m)$. These signatures are commonly understood to be generated with the help of an Elliptic Curve cryptosystem, such one of the EC-DSA schemes standardized in [2].

### B. Hybrid Scheme

The Hybrid scheme, proposed in [5], is essentially a pseudonymous authentication scheme that enables nodes to generate their own certified pseudonyms. The only aspect of BP that Hybrid changes is the computation of pseudonym and its certificate and, consequently, their validation. To maintain the degree of privacy protection pseudonymous authentication provides, nodes utilize a group signature (GS) scheme [4] to generate their own certificates.

In brief, a GS scheme requires that each node $V$ is equipped with a secret *group signing key* $gsk_V$ and that it belongs to a *group* $G$, with members all vehicles registered with the CA. A *group signature* $\Sigma_{CA,V}$ generated by a group member can be validated with the use of the *group public key* $gpk_{CA}$. The essence of a GS scheme is that any $V$ can sign a message on behalf of the group, but the identity of $V$ is never revealed to any signature verifier and no two signatures of a legitimate group member can be linked.

The GS scheme allows each node $V$ in $G$ to first generate its own set of pseudonyms $\{K_V^i\}$ (and their corresponding private keys $k_V^i$) for a "classic" cryptosystem such as EC-DSA. Then, $V$ generates a group signature $\Sigma_{CA,V}()$ on each pseudonym $K_V^i$. What $V$ does is to "self-certify" $K_V^i$ by producing $Cert_{CA}^H(K_V^i)$; $H$ denotes the hybrid scheme, to differentiate this from the certificate of the BP approach, and $CA$ indicates that the certificate was generated by a legitimate node registered with the CA (the group G).

When a Hybrid-signed message is received, the group signature $\Sigma_{CA,V}(K_V^i)$ is first validated. According to the GS scheme properties, the verifier of the certificate *cannot* identify $V$ and *cannot* link this certificate and pseudonym to any prior pseudonym used by $V$. Once the certificate is validated, the "classic" signature $\sigma_{k_V^i}(m)$ identical to that of the BP scheme is validated. For further details, omitted here due to space limitations, we refer to [5] and references within.

### C. Optimizations

We consider the three optimizations devised in [5], but when applicable (for Optimizations 2 and 3 and Optimization 1 at the verifier's side) we do so for any pseudonymous authentication scheme, rather than the Hybrid scheme only. When the optimizations are applicable for both the BP and Hybrid schemes, we simplify the notation for the certificates to $Cert(K_V^i)$, not distinguishing which method is used for the certificate generation.

*Optimization 1:* At the sender side, the certificate $Cert_{CA}^H(K_V^i)$ for a pseudonym $K_V^i$ is computed only once, as $Cert_{CA}^H(K_V^i)$ remains unchanged throughout the *pseudonym lifetime* $\tau$. Similarly, at the verifier, $Cert(K_V^i)$ is validated (and stored) only when it is first received, although the same signer will append it to multiple subsequent messages. Optimization 1 is effective as the *beaconing rate* $\gamma$ is such that $\tau >> \gamma^{-1}$.

*Optimization 2:* The sender appends its signature $\sigma_{k_V^i}(m)$ to all messages, but it appends it along with the corresponding $K_V^i, Cert(K_V^i)$ once every $\alpha$ messages (beacons). We denote such a message as $LONG$. For the remaining $\alpha - 1$ messages, $V$ appends only $\sigma_{k_V^i}(m)$ to the message payload. We denote such messages as $SHORT$. We term $\alpha$ the *Certificate Period*. Once a new pseudonym must be used (e.g., upon expiration of the one currently in use by $V$), $\sigma_{k_V^{i+1}}(m), K_V^{i+1}, Cert(K_V^{i+1})$ is transmitted.

*Optimization 3:* As Optimization 2 can harm the protocol robustness,[1] the transmission of $K_V^{i+1}, Cert(K_V^{i+1})$ can be repeated for $\beta$ consecutive messages after the pseudonym change, with $\beta$ termed the *Push Period*.

---

[1]If the message with $K_V^{i+1}, Cert(K_V^{i+1})$ is lost due to wireless channel impairments, nodes in range of $V$ will be able to validate any signed safety messages only $\alpha$ (or more) messages after the lost pseudonym/certifcate transmission.

## III. SIMULATION SETUP

Our simulator, custom-built in C and partly relying on ns-2, implements the *Dedicated Short Range Communications (DSRC)* data link [1], with vehicles transmitting *beacons* on a common channel at a rate $\gamma = 10$ beacons/sec, a value considered mandatory for safety applications. Vehicles include their direction and location but are not further concerned with the content of messages $m$ whose payload is set to 200 bytes. Table II provides the cryptographic communication overhead per message. We implement the physical layer and a realistic radio propagation model [10], as in [6], [12], with an intended (nominal) transmission range of $d = 200$ meters [1], [13].

We simulate scenarios of four- and eight-lane highways,[2] with vehicles moving in two opposing two- and four-lane flows. Along each lane, vehicles are randomly placed with an average spacing of $s$ meters between two vehicles, and their velocity is randomly drawn with an average $v$. The average spacing value is one vehicle every $s = 20$ meters per lane, and the average vehicle velocity is $v = 80 km/h$. Vehicles process messages from vehicles with the same heading within a $\delta > 0$. We model wet road conditions by setting the vehicle braking capabilities to 4 $m/s^2$, which correspond to a friction coefficient $\mu_k = 0.41$.

We consider an "emergency braking alarm" application, with such messages transmitted independently of the beaconing pattern. We investigate how many vehicle collisions occur when safety messaging is used with and without security. We focus on a platoon of one hundred cars, denoted as $V_1$ to $V_{100}$, along a single lane, with $V_1$ at the head and $V_{100}$ at the queue of the platoon. The leading vehicle $V_1$ makes an emergency brake and starts sending warning messages. Once some $V_i$, with $i > 1$ receives the warning, it warns its driver and starts sending warnings itself. As proposed in [3], [7], [14], when $V_i$ receives a warning from a $V_j$ with $j > i$, it stops transmitting warnings. This approach assumes that at least one vehicle behind $V_i$ has already been warned, thus $V_j$ is in a better position to keep warning $V_k$, for $k > j$. We assume that all braking actions here are (and reported as) emergency braking.

Braking has two effects: (i) it turns on the vehicle rear red lights that warn visually vehicles within range (and line) of sight (expressed in meters, depending on the simulated weather conditions), and (ii) it triggers the transmission of own warning messages. Besides warning other vehicles, a warned $V_i$ clearly warns its driver who starts braking shortly after. Driver reactions, modeled as a random reaction delay between 0.75 and 1.5 seconds, are triggered by both VC-enabled and visual (red light) warnings.

These simulated conditions are particularly challenging, with high vehicle density and average velocities that are high for the given average vehicle spacing. High vehicle density results in high network load, especially given the high beaconing rates, and thus increased packet loss rates. We do not consider any optimization to reduce processing overhead (e.g., based on the message content or by having vehicles adapt their beaconing rates), beyond considering messages from vehicles with the same heading. This is reasonable for shoulder-separated highways, but it hard to achieve in an urban or rural setting. In the latter case, of course, it would be unrealistic to consider densities of, for example, an average 160 vehicles within a vehicles's nominal range, which could be the case in an eight-lane highway with an average vehicle spacing per lane of 20 meters. Finally, we remark that we do not simulate lane-changing as a means to avoid collisions, but leave it as an item for future work. Depending on road conditions (e.g., in low vehicle density) lane changing may lead to fewer collisions, but it could have the same effect across all scenarios (with or without security).

We choose pseudonym lifetime $\tau = 60$ sec. We consider the first 60 sec of the simulated time as a warm-up period, during which no emergency conditions arise. By the end of the warm up period, this allows for one pseudonym change and discovery of neighbors by validating their corresponding certificates. This is a realistic situation: at any point in time, thus when an emergency arises, vehicles have already discovered some of their neighbors and can immediately validate their warnings. The simulation concludes when all vehicles in the platoon are immobile;, as $V_1$ does not resume any motion after its emergency braking.

## IV. COMMUNICATION OVERHEAD

Our results are shown in Figs. 1 and 2 respectively: the y axis is the probability of successful reception of a packet, as a function of the distance between the packet sender and receiver (the x-axis). We emphasize that although the message processing load may be trimmed down, e.g., by using the node mobility dynamics, this is not the case for the channel load: any transmission 'contributes' interference and increases the likelihood of packet collisions.

This is why the probability of message reception is overall lower in Fig. 2 than in Fig. 1. The increase from four to eight lanes essentially doubles vehicle density and network load. Looking at each plot individually, safety messaging with "No Security" achieves the highest reliability consistently as the distance changes. In contrast, the BP or Hybrid with $\alpha = 1$ achieves the lowest reliability. This due to transmission overhead: "No Security" means a constant packet size of 200 bytes, while each of the different $\alpha$ settings corresponds to different, higher per-packet overhead, as shown in Table I. Thanks to Optimization 2, communication reliability improves as $\alpha$ increases, but more so for Hybrid whose certificates are larger. The average communication overhead due to cryptography is calculated in Table I for the Hybrid and BP schemes, as a function of $\alpha$. "Pushing" the Optimization 2 beyond a certain point does not significantly improve reliability further.

## V. PROCESSING OVERHEAD

We measure the number of packets that a given receiver, $\mathcal{R}$, must process per time unit. The processing overhead depends

---

[2]We also conducted experiments with a six-lane highway omitted here due to space limitations.

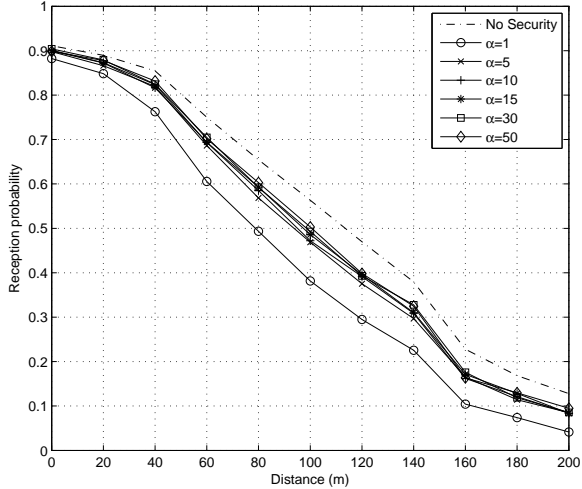

Fig. 1. Probability of Successful Packet Reception; Highway Scenario, Baseline Pseudonym scheme (BP), 8 lanes.

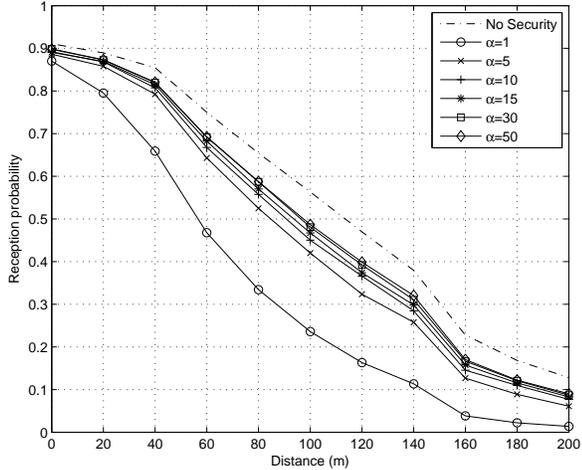

Fig. 2. Probability of Successful Packet Reception; Highway Scenario, Hybrid scheme, 8 lanes.

| $\alpha$ | 1 | 5 | 10 | 15 | 30 | 50 |
|---|---|---|---|---|---|---|
| **BP** | 341 | 266 | 257 | 254 | 251 | 250 |
| **Hybrid** | 502 | 299 | 273 | 265 | 257 | 253 |

TABLE I

AVERAGE PACKET SIZE, INCLUDING OVERHEAD, IN BYTES, FOR BP AND HYBRID AND DIFFERENT $\alpha$ VALUES.

|  | Sign (ms) | Verify (ms) | Overhead (bytes) |
|---|---|---|---|
| **BP** *LONG* | 1.3 | 7.2 | 141 |
| **Hybrid** *LONG* | 54.2 | 52.3 | 302 |
| *SHORT* | 0.5 | 3 | 48 |

TABLE II

PROCESSING COSTS AND OVERHEAD FOR DIFFERENT PACKET TYPES, GIVEN THE BENCHMARKS IN [5].

on the rate of messaging *received* at $\mathcal{R}$, and then the employed security scheme that specifies at what rate (in fact, how many or those received) must be *processed*. For simplicity, we consider as time unit one *beacon period*, i.e., $\gamma^{-1}$ seconds, which we refer to as one *slot*. On-board hardware platforms that will be used for commercial deployment of vehicular communication systems are not known and neither is their expected processing power. This is why we refrain from attempting to answer if $\mathcal{R}$ would be able to process all the messages it should. Instead, we measure the rate of messages $\mathcal{R}$ should process.

Then, only as a reference, we consider platforms with capabilities close to those used as reference in [5]. We summarize the per-message costs in Table II, and the resultant maximum number of messages that could then be processed in one slot in Table III, assuming all messages are of the same type and the node performs no other proessing. Only verifications are considered because they are dominant factor of cryptographic processing overhead. With $n$ relevant senders within range, $\mathcal{R}$ would receive and process roughly $n$ messages per slot, versus one signature generation. The security schemes in Sec. II involve two types of messages, one carrying a signature only, termed the $SHORT$, and one carrying the signature and the pseudonym and the certificate of the signer, termed the $LONG$ message. One $LONG$ message is transmitted every $\alpha$ $SHORT$ messages, and $\beta$ consecutive $LONG$ messages are transmitted upon a pseudonym change.

Considering that these are all safety messages, $\mathcal{R}$ must in principle process all of them. Those are, as explained above, roughly the messages originating from the $n$ vehicles traveling in the same direction as $\mathcal{R}$. This arrival rate of messages at $\mathcal{R}$ depends on the number of relevant neighbors $n$, at any given time $t$, the message generation rate $\gamma$, the type of messages generated, and the radio propagation.

The node is able to process all the messages received within the slot, as the safety application mandates, if the total processing time is less than the slot duration: if $\sum_{i=0}^{n} t_i < \gamma^{-1}$, where $t_i$ is the time needed to process the $i$-th message. The rate $\mu$ at which messages should be processed by $\mathcal{R}$ should basically equal that of relevant arriving messages, minus some messages that the Optimizations allow avoiding to process. Not all received $LONG$ messages from a given sender $V_i$ must be processed; the first one for each $V_i$ suffices to validate the pseudonym in use. Conversely, $SHORT$ messages from

|  | **Packets per beacon period (slot) of $\gamma^{-1}$ sec** |
|---|---|
| **BP** $LONG$ | 13.9 |
| **Hybrid** $LONG$ | 1.9 |
| $SHORT$ | 33.3 |

TABLE III

MAXIMUM NUMBER OF PACKETS PER TIME SLOT THAT CAN BE VERIFIED.

| $\alpha$ | LONG | | | | SHORT | | | |
|---|---|---|---|---|---|---|---|---|
| | Received | | Processed | | Received | | Processed | |
| | $\mu_R$ | $\sigma_R$ | $\mu_P$ | $\sigma_P$ | $\mu_R$ | $\sigma_R$ | $\mu_P$ | $\sigma_P$ |
| 1 | 22.73 | 2.69 | 0.07 | 0.27 | 0.00 | 0.00 | 0.00 | 0.00 |
| 5 | 5.40 | 1.52 | 0.07 | 0.26 | 20.46 | 2.40 | 20.44 | 2.40 |
| 10 | 2.94 | 1.32 | 0.07 | 0.26 | 23.56 | 2.55 | 23.54 | 2.57 |
| 15 | 2.06 | 1.37 | 0.07 | 0.25 | 24.66 | 2.61 | 24.66 | 2.61 |
| 30 | 1.16 | 1.10 | 0.07 | 0.25 | 25.77 | 2.60 | 25.77 | 2.60 |
| 50 | 0.83 | 0.87 | 0.07 | 0.26 | 26.07 | 2.72 | 26.03 | 2.72 |

TABLE IV

PACKET RECEPTION AND PROCESSING STATISTICS FOR A SCENARIO OF 4 LANES OF VEHICLES AND OPTIMIZATION 3 IN USE - MEASURED RESULTS FOR HYBRID.

| $\alpha$ | LONG | | | | SHORT | | | |
|---|---|---|---|---|---|---|---|---|
| | Received | | Processed | | Received | | Processed | |
| | $\mu_R$ | $\sigma_R$ | $\mu_P$ | $\sigma_P$ | $\mu_R$ | $\sigma_R$ | $\mu_P$ | $\sigma_P$ |
| 1 | 25.67 | 3.30 | 0.14 | 0.37 | 0.00 | 0.00 | 0.00 | 0.00 |
| 5 | 7.52 | 2.69 | 0.13 | 0.34 | 28.63 | 3.97 | 28.56 | 3.97 |
| 10 | 4.24 | 1.95 | 0.13 | 0.36 | 33.70 | 3.74 | 33.56 | 3.71 |
| 15 | 2.97 | 1.32 | 0.13 | 0.36 | 35.36 | 3.74 | 34.99 | 3.72 |
| 30 | 1.68 | 1.23 | 0.14 | 0.35 | 37.62 | 3.80 | 37.33 | 3.74 |
| 50 | 1.20 | 1.06 | 0.13 | 0.36 | 38.37 | 4.15 | 37.87 | 4.06 |

TABLE V

PACKET RECEPTION AND PROCESSING STATISTICS FOR A SCENARIO OF 8 LANES OF VEHICLES AND OPTIMIZATION 3 IN USE - MEASURED RESULTS FOR HYBRID.

a non-yet-validated $V_i$ do not have to be processed.

We perform a series of simulations for the Hybrid scheme for which the packet reception probability was determined in Sec. IV. We performed the same experiments for the BP scheme, but we omit those results due to space limitations and because the difference between $LONG$ and $SHORT$ packets are more pronounced for the Hybrid. Table IV provides the results for a scenario with four lanes of traffic and 80 vehicles on the average, with the messages originating from roughly 40 of them being of interest to $\mathcal{R}$. Table V corresponds to the case of 8 lanes of traffic and 160 vehicles on the average within range of $\mathcal{R}$ (similarly, with approximately 80 of them generating messages that would be processed by $\mathcal{R}$). The measured numbers in both tables, obtained when $\beta = 5$ for both settings, are the means and standard deviations of received packets per slot, $\mu_R$ and $\sigma_R$ respectively, and the means and standard deviations of processed packets per slot, $\mu_P$ and $\sigma_P$ respectively. The results are shown in these tables for $SHORT$ and $LONG$ packets separately, as a function of $\alpha$.

Note that since there is no specific processing power for $\mathcal{R}$, the measured statistics for processed messages are both the values for the rates that would need to (and are then trivially in this case) achieved by $\mathcal{R}$. For $\alpha = 1$, all traffic is composed solely of $LONG$ packets, but clearly this is not an advisable setting neither for the BP nor the Hybrid scheme, as the results in Secs. IV and VI show. As a side-note, given the processing capabilities in Table III, we can deduce that a GS scheme [5] would not be workable for the considered platform, as all received $LONG$ messages (in fact, at a somewhat higher rate than that measured in our tables here, as a GS alone results in lower overhead) would need to be processed at $\mathcal{R}$.

Returning to the Hybrid scheme, only one $LONG$ message per $\tau$, the pseudonym lifetime, and per vehicle (sender) must be processed.[3] This overhead can be easily undertaken, as validation of *processed LONG* messages (e.g., 0.14 per slot) would require approximately 8% of the processing resources reported in Table III. The high cost for a single Hybrid $LONG$ message is amortized.

It is interesting however that the $SHORT$ message pro-cessing becomes a dominant load factor. In Table IV, $\mu_P$ could be supported by the considered platform, but it would consume 78% of its available power. If we consider the highest measured mean plus three times the standard deviation, the resultant $SHORT$ processed rate combined with a similar estimate of the worst-case $LONG$ rate could not be sustained; there would be slots in which arrivals exceed $\mathcal{R}$'s processing capability. This is certainly so for the 8-lane scenario, as shown in Table V: only $SHORT$ messages exceed the 33.3 maximum achievable processing rate (Table III). These observations call for action on two fronts, (i) an increase of the processing power and (ii) the design of a set of optimizations in order to avoid processing all the packets arriving at $\mathcal{R}$.

---

[3]The same is true for $BP$ with Optimizations.

## VI. TRANSPORTATION SAFETY

We simulate the impact of safety messaging without security, denoted as "No Security", versus that of the security protocols without Optimization 3, i.e., $\beta = 0$, and that of the protocols with Optimization 3 enabled with a $\beta = 5$. We assume that all vehicles can process all messages needed within a slot (with increased processing power, as explained in Sec. V). Figs. 3, 4 plot the percentage of crashed platoon vehicles as a function of $\alpha$.

At first, we observe that safety messaging reduces crashes to around 10% of the platoon vehicles, while in the absence of vehicle communications, for exactly the same scenarios, 80%-100% of the vehicles crash (not shown in the figures). Recall that the simulated scenarios are challenging (e.g., dense placement of fast-moving vehicles). "No Security" safety messaging (which reduces crashes up to 90% with respect to no V2V) achieves the lowest percentage of crashes and is used as a benchmark; $\alpha$ is not a relevant parameter and curves in Figs. 3, 4 are flat with minor variability due to the randomly seeded simulations.

"No Security" safety messaging is the most effective due to the lowest network overhead and no restrictions on which alert message can be validated. In contrast, the tuning of the secure VC protocols affects the effectiveness of the safety application. In Figs. 3, 4, we observe at first fewer crashes as $\alpha$ increases, but then a slight increase in crashes for high $\alpha$ values. On the one hand, the increase of $\alpha$ reduces the channel load and thus increases the per-packet reception probability. On the other hand, when $\alpha$ takes high values, the authentication delay for a receiver missing a $LONG$ message increases: for example, for $\alpha = 50$ the authentication delay after the loss of one $LONG$ message is at least 5 seconds.

Optimization 3 reduces crashes for a given $\alpha$ compared to the non-optimized protocol ($\beta = 0$), adding low overhead but also significantly reducing authentication delays and averting situations that would otherwise lead to crashes. Figs. 3, 4 show that $\beta = 5$ yields an improvement exactly for high $\alpha$ values. From a different point of view, Optimization 3 brings the effectiveness of the application close to that of "No Security". Overall, the investigated safety application on top of the Hybrid, with appropriate tuning, is equally effective to that on top of the BP.

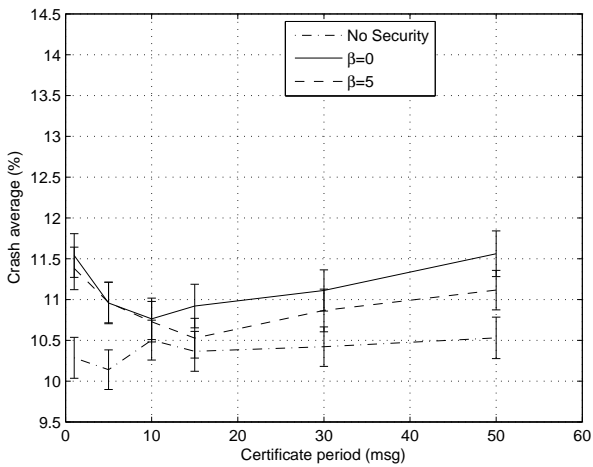

Fig. 3. Collisions in a Platoon of 100 Cars; 8-lane Highway Scenario, Baseline Pseudonym Scheme

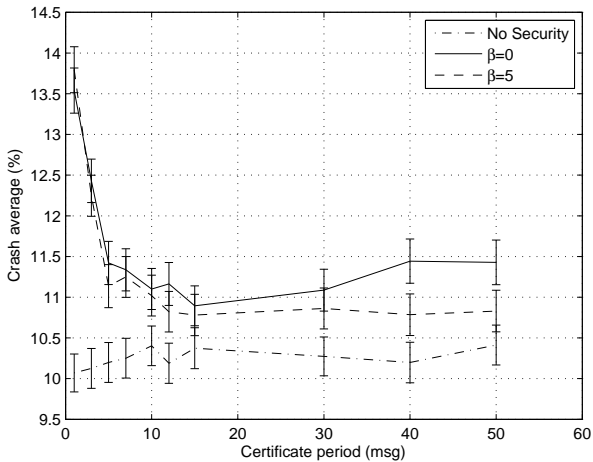

Fig. 4. Collisions in a Platoon of 100 Cars; 8-lane Highway Scenario, Hybrid Scheme

## VII. CONCLUSIONS

We analyzed through simulations the cost and impact on transportation safety that security and privacy-enhancing technologies in the literature have. We obtained packet reception probability curves for each security scheme, measured the rates of packets each node must process, and developed a simulation environment that adapts vehicle mobility according to received safety application messages to analyze the safety messaging and security impact on transportation safety.

We found that current security protocols push on-board processors considered nowadays to their limits, or above them, calling for either a redesign of the VC protocols or increased on-board processing power. If the latter is available, secure VC schemes can enable safety levels for an emergency braking application nearly identical to those achieved without security. Beyond these encouraging results, further analysis and characterization of the processing overhead at each node, along with investigations of alternative communication protocols, as well as field experimentation are needed as future work.

## REFERENCES

[1] DSRC: Dedicated short range communications. http://grouper.ieee.org/groups/scc32/dsrc/index.html.
[2] IEEE 1363a 2004. IEEE standard specifications for public-key cryptography- amendment 1: Additional techniques, 2004.
[3] S. Biswas, R. Tatchikou, and F. Dion. Vehicle-to-vehicle wireless communication protocols for enhancing highway traffic safety. *IEEE Communications Magazine*, 44:74– 82, January 2006.
[4] D. Boneh and H. Shacham. Group signatures with verifier-local revocation. In *CCS '04*, pages 168–177, Washington DC, USA, October 2004.
[5] G. Calandriello, P. Papadimitratos, J.-P. Hubaux, and A. Lioy. Efficient and Robust Pseudonymous Authentication in VANET. In *Proceedings of the Fourth ACM International Workshop on Vehicular Ad hoc Networks (VANET)*, pages 19–28, Montreal, QC, Canada, September 2007.
[6] Q. Chen, F. Schmidt-Eisenlohr, D. Jiang, M. Torrent-Moreno, L. Delgrossi, and H. Hartenstein. Overhaul of IEEE 802.11 modeling and simulation in ns-2. In *MSWiM '07: Proceedings of the 10th ACM Symposium on Modeling, analysis, and simulation of wireless and mobile systems*, pages 159–168, Chania, Crete, Greece, 2007.
[7] T. ElBatt, S. K. Goel, G. Holland, H. Krishnan, and J. Parikh. Co-operative collision warning using dedicated short range wireless communications. In *Proceedings of the Third international workshop on Vehicular Ad hoc Networks (VANET)*, pages 1–9, Los Angeles, CA, USA, September 2006.
[8] M. Gerlach, A. Festag, T. Leinmller, G. Goldacker, and C. Harsch. Security architecture for vehicular communication. In *WIT 2005*, Hamburg, Germany, 2007.
[9] IEEE1609.2. IEEE trial-use standard for wireless access in vehicular environments - security services for applications and management messages, July 2006.
[10] M. Nakagami. The m-distribution, a general formula of intensity distribution of the rapid fading. In W. G. Hoffman, editor, *Statistical methods in radio wave propagation*, pages 3–36, Oxford, 1960. Pergamon.
[11] P. Papadimitratos, L. Buttyan, J-P. Hubaux, F. Kargl, A. Kung, and M. Raya. Architecture for Secure and Private Vehicular Communications. In *Proceedings of the Seventh International Conference on ITS Telecommunications (ITST)*, pages 1–6, Sophia-Antipolis, France, June 2007.
[12] M. Torrent-Moreno, D. Jiang, and H. Hartenstein. Broadcast reception rates and effects of priority access in 802.11-based vehicular ad-hoc networks. In *VANET '04*, pages 10–18, Philadelphia, PA, USA, October 2004.
[13] Q. Xu, T. Mak, J. Ko, and R. Sengupta. Vehicle-to-vehicle safety messaging in DSRC. In *VANET '04*, pages 19–28, Philadelphia, PA, USA, October 2004.

[14] X. Yang, J. Liu, F. Zhao, and N. H. Vaidya. A vehicle-to-vehicle communication protocol for cooperative collision warning. In *First Annual International Conference on Mobile and Ubiquitous Systems: Networking and Services (MobiQuitous'04)*, 2004.